# Energy-Aware Task Partitioning on Heterogeneous Multiprocessor Platforms

Elsayed Saad[1], Medhat Awadalla[1, 3], Mohamed Shalan[2] and Abdullah Elewi[1]

[1] Electronics, Communication and Computer Engineering Department, Helwan University
Cairo, Egypt

[2] Computer Science and Engineering Department, the American University in Cairo
Cairo, Egypt

[3] Electrical and Computer Engineering Department, Sultan Qaboos University
Muscat, Oman

**Abstract**
Efficient task partitioning plays a crucial role in achieving high performance at multiprocessor platforms. This paper addresses the problem of energy-aware static partitioning of periodic real-time tasks on heterogeneous multiprocessor platforms. A Particle Swarm Optimization variant based on Min-min technique for task partitioning is proposed. The proposed approach aims to minimize the overall energy consumption, meanwhile avoid deadline violations. An energy-aware cost function is proposed to be considered in the proposed approach. Extensive simulations and comparisons are conducted in order to validate the effectiveness of the proposed technique. The achieved results demonstrate that the proposed partitioning scheme significantly surpasses previous approaches in terms of both number of iterations and energy savings.

## 1. Introduction

Nowadays, embedded systems are involved in most details of our life such as smart phones, pocket PCs, Personal Digital Assistants (PDAs), multimedia devices, ... etc. As the applications on these devices are being complicated, there is a need to increase the performance while keeping the energy consumption of these devices in accepted levels especially for the portable battery-powered ones. So, minimizing energy consumption to prolong the battery life while achieving higher performance is a critical issue in the design of portable embedded systems.

As the processor is one of the most important power consumers in any computing system, today's chip multiprocessor (CMP) or multiprocessor system on chip (MPSoC) platforms can deliver a higher performance at the cost of lower power consumption than uniprocessor systems.

Embedded systems today are often implemented upon platforms comprised of different kinds of processing units, such as CPU's, DSP chips, graphics co-processors, math co-processors, etc., with each kind of processing unit specialized to perform a different function most efficiently. Such platforms are commonly referred to as heterogeneous platforms [1]. TI's OMAP™ [2] mobile processors are good example of these heterogeneous platforms.

The multiprocessor scheduling of recurrent real-time tasks can be generally carried out under the partitioned scheme or under the global scheme. In the partitioned scheme, the tasks are statically partitioned among the processors and all instances (jobs) of a task are executed on the same processor and no job is permitted to migrate among processors. In the global scheme, a task can migrate from one processor to another during the execution of different jobs. Furthermore, an individual job of a task that is preempted from some processor, may resume execution in a different processor. Nevertheless, in both schemes, parallelism is prohibited, i.e., no job of any task can be executed at the same time on more than one processor.

This paper considers the partitioned scheduling scheme. The main advantage of the partitioned scheduling is that after partitioning the tasks among processors, the multiprocessor scheduling problem is reduced to a set of traditional uniprocessor ones.

The problem of partitioning tasks among processors, sometimes [3, 8] referred to as Task Assignment Problem (TAP), is an intractable NP-Hard problem [1] even if the processors are homogeneous [4]. So, approximation algorithms and heuristic techniques are used to solve this problem. This paper proposes a Particle Swarm

Optimization (PSO) variant based on Min-min technique for energy-aware task partitioning on heterogeneous multiprocessor platforms.

The rest of this paper is organized as follows: Section 2 reviews existing research on task partitioning upon heterogeneous platforms and related areas. Section 3 defines the problem and describes task, processor, and power models used in this paper. Section 4 describes PSO and Min-min techniques for task partitioning and introduces our proposed approach. Section 5 presents simulation results for the proposed algorithm and discusses these results. Section 6 summarizes our conclusions.

## 2. Related Work

Baruah [1] proved that task partitioning among heterogeneous multiprocessors is intractable (strongly NP-hard), represented the problem as an equivalent Integer Linear Programming (ILP) problem, and designed a 2-step approximation algorithm for solving this problem. The idea of LP relaxations to ILP problems is used in the first step to map most tasks, while in the second step the algorithm maps the remaining tasks using exhaustive enumeration. This two-step algorithm takes time polynomial in the number of tasks, and exponential in the number of processors. The same author [5] then used tree-partitioning in the second step instead of exhaustive enumeration to make the algorithm takes time polynomial in the number of tasks, and polynomial in the number of processors.

In [6], Braun et. al. compared 11 heuristics for mapping a set of independent tasks onto heterogeneous distributed computing systems. The best one that has minimum makespan, that is defined as the maximum completion time for the whole processors, was the Genetic Algorithm (GA) followed by Min-min algorithm.

Chen and Cheng [7] applied the Ant Colony Optimization (ACO) algorithm. They proved that ACO outperforms both GA and LP-based approaches in terms of obtaining feasible solutions as well as processing time.

In [3], Abdelhalim presented a modified algorithm based on the Particle Swarm Optimization (PSO) for solving this problem and showed that his approach outperforms the major existing methods such as GA and ACO methods. Then, his PSO approach is developed to can further optimize the solution to reduce the energy consumption by minimizing average utilization of processors (without using any energy or power model). Finally, a tradeoff between minimizing the design makespan as well as energy consumption is obtained.

Visalakshi and Sivanandam [8] presented a hybrid PSO method for solving the task assignment problem. Their algorithm has been developed to dynamically schedule heterogeneous tasks onto heterogeneous processors in a distributed setup. It considers load balancing and handles independent non-preemptive tasks. The hybrid PSO yields a better result than the normal PSO when applied to the task assignment problem. The results are also compared with GA. The results infer that the PSO performs better than the GA.

In [9], Omidi and Rahmani used PSO for task scheduling in multiprocessor systems as an important step for efficient utilization of resources. They considered independent tasks on homogeneous multiprocessor systems.

Apart from all these efforts, this paper integrates the PSO approach with a polynomial-time partitioning technique; Min-min. The proposed approach takes into account energy efficiency during task partitioning among heterogeneous cores in MPSoCs.

## 3. System Model

This paper considers the problem of power-aware task partitioning on heterogeneous multiprocessor platforms. So, models of task, processor, and power are presented.

3.1 Task Model

A periodic real-time task $\tau_i$ generates an infinite sequence of task instances (jobs). Each job executes for $C$ time units at most, be generated every $T$ time units, and has a relative deadline $D$ time units after its arrival.

This paper considers a periodic task set $\{\tau_1, \tau_2, ..., \tau_n\}$ of $n$ independent real-time tasks. A task $\tau_i$ is represented as 3-tuple $\tau_i = (C_{ij}, D_i, T_i)$ where $C_{i,j}$ is the Worst-Case Execution Time (WCET) of task $\tau_i$ on processor $j$, $D$ is the relative deadline, and $T$ is the period. Implicit deadlines are considered in this paper, i.e., the relative deadline is assumed to be the same as the period. Each task $\tau_i$ has a utilization $u_{ij} = C_{ij} / T_i$ on processor $j$.

An $n \times m$ utilization matrix as in [1] can be defined where each row represents a task and each column represents a processor.

## 3.2 Processor Model

A heterogeneous multiprocessor platform with $m$ preemptive processors based on CMOS technology is defined as $\{P_1, P_2, ..., P_m\}$.

This paper considers Dynamic Voltage/Frequency Scaling (DVFS) processors that supports variable frequency (speed) and voltage levels continuously, i.e., DVFS processors can operate at any speed/voltage in its range (ideal). Of course, practical DVFS processors supports discrete speed/voltage levels (non ideal). So, the desired speed/voltage of the ideal DVFS processor is rounded to the nearest higher speed/voltage level the practical DVFS processor supports.

The time (energy) required to change the processor speed is very small compared to that required to complete a task. It is assumed that the speed/voltage change overhead, similar to the context switch overhead, is incorporated in the task execution time. In this work, it is assumed that the processor's maximum speed (frequency) is 1 and all other speeds are normalized with respect to the maximum speed.

When MPSoCs platforms are considered, there are the per-core and full-chip DVFS techniques [10]. In the per-core DVFS, each core operates at individual frequency/voltage, and has no operating frequency constraint. On the other hand, the practical full-chip DVFS designs restrict that all the cores in one chip operate at the same clock frequency/voltage.

For each processor, the tasks are scheduled according to Earliest Deadline First (EDF) scheduling algorithm. So, a processor utilization $U_j$ which is the sum of the utilizations of tasks assigned to this processor can not exceed 1, i.e., $U_j = \sum_i u_{ij} \leq 1$.

## 3.3 Power Model

The power consumption in CMOS circuits has two main components: dynamic and static power. The dynamic power consumption which arises due to switching activity can be represented as [11]:

$$P_{dynamic} = C_{eff}.V_{dd}^2.f \qquad (1)$$

Where $C_{eff}$ is the effective switching capacitance, $V_{dd}$ is the supply voltage, and $f$ is the processor clock frequency (speed) which can be expressed in terms of a constant $k$, supply voltage $V_{dd}$ and threshold voltage $V_{th}$ as follows:

$$f = k.(V_{dd} - V_{th})^2 / V_{dd} \qquad (2)$$

The static power consumption is primarily occurred due to leakage currents ($I_{leak}$) [12], and the static (leakage) power ($P_{leak}$) can be expressed as:

$$P_{leak} = I_{leak}.V_{dd} \qquad (3)$$

When the processor is idle, a major portion of the power consumption comes from the leakage. Currently, leakage power is rapidly becoming the dominant source of power consumption in circuits and persists whether a computer is active or idle [12].
So, lowering supply voltage is one of the most effective ways to reduce both dynamic and leakage power consumption. As a result, it reduces energy consumption where the energy consumption is the power dissipated over time.

For simplicity reasons, Eq. (1) is reduced to a simplified power model $P = f^3$ using normalized values where $f$ is the processor speed (frequency). Then, a simplified energy model $E = f^2$ (using normalized values) can be used.

## 4. The Proposed Approach

Before introducing our proposed approach in this paper, a background on PSO and Min-min techniques will be presented.

### 4.1 PSO

Kennedy and Eberhart [13] developed the PSO algorithm simulating the behavior of swarms in the nature, such as birds, fish, etc. In PSO, the potential solutions, called particles, fly through the problem space by following the current optimum particles. PSO has been successfully applied in many scientific areas and there are many variants of the algorithm. A survey of PSO methods and applications could be found in [14].

At the beginning, a set (swarm) of random solutions (particles) is used to initialize the PSO algorithm that starts iterations looking for optimal solution. During every iteration, each particle is updated by two best values. The first one is the personal best *pbest* that the particle has achieved so far. The second is the global best *gbest* obtained by any particle in the swarm. After finding the two best values, the particle updates its velocity and position according to equations (4) and (5) respectively. The typical procedure of PSO is shown in figure 1.

```
initialize the population randomly.
DO
{
   For each particle.
   {
      Calculate fitness value If the fitness value is better than the best
      fitness value (pbest) in history then set current value as the new
      pbest.
   }
   Choose the particle with the best fitness value of all particles as
   the gbest.
   For each particle.
   {
      Calculate new velocity:

      V_new = W.V_old + C_1.R_1.(pbest - X) + C_2.R_2.(gbest-X)     (4)

      (Where W is inertia constant, R_1 and R_2 are random values. C_1
      and C_2 are constant values and X is particle position.)
      Update particle position:

      X_new = X_old + V_new                                          (5)
   }
}
Until termination criterion is met.
```

Fig. 1 The typical procedure of PSO [9].

The random numbers $R_1$ and $R_2$ are generated uniformly between 0 and 1 and the constants $C_1$ (self-knowledge factor) and $C_2$ (social-knowledge factor) are usually in the range from 1.5 to 2.5. Finally, the inertia factor $W$ can be fixed or varied with a decreasing value as the algorithm proceeds [8] or it may be restarted as in [3].

PSO has been applied to solve the problem of task partitioning for homogeneous multiprocessor as in [9] and also for heterogeneous multiprocessors [3, 8].

Considering a system consisting of $m$ processors and $n$ tasks. A possible solution (particle) is a vector of $n$ elements, where each element is associated to a given task. Each element takes an integer value $i$ where $1 \leq i \leq m$ and represents the processor that the task is assigned to.
Thus, the search space size is $m^n$
There are $k$ particles in the swarm that form swarm (population) size; these particles are initialized randomly.

### 4.2 Min-min

The Min-min [6] algorithm is originally designed for mapping tasks in heterogeneous computing systems and does not consider real-time tasks. It first finds the minimum completion time of all unmapped tasks, where the completion time of a task on a machine equals task's execution time on that machine plus execution times of all tasks mapped to that machine. Next, the task which has minimum completion time is selected, similar technique called Max-min selects the task with maximum completion time, and mapped to the machine. Finally, the newly mapped task is removed and the process repeats until all tasks are mapped.

To handle real-time tasks on multiprocessor system, task utilization is considered instead of execution time and completion utilization is used. Of course, tasks that make the processor's utilization exceeds 1 are unaccepted. If there is no accepted alternative, then the task set is unfeasible.

### 4.3 The Proposed Min-min based PSO Approach

The Min-min based PSO approach, proposed in this paper, simply modifies the initialization step in the PSO procedure by incorporating a Min-min solution (particle) in the randomly generated population. This approach gives the PSO algorithm a push to start from a good solution and then the PSO goes on trying to optimize the solution resulting in the Min-min solution in the worst case.

Firstly, A cost function favoring makespan (maximum processor accumulative utilization) minimization is proposed. Then, a penalty is added to the infeasible solutions that exceed the processing capacity of any processor. In other words, the cost is represented as follows [3]:

$Cost = Max(U_j) + Penalty$      for $j = 1,2, .. , m$     (6)

$Penalty = Sum(U_j > 1)$         for $j = 1,2, .. , m$     (7)

Next, the cost function is developed to incorporate energy where the proposed PSO approach tries to find energy-efficient solutions. Aydin and Yang [4] considered energy-aware task partitioning for homogeneous multiprocessors and introduced some helpful proofed theorems and propositions. Some of them are presented here.

**Proposition 1** [4] *For a single processor system and a set of periodic real-time tasks with total utilization ≤ 1. The optimal speed to minimize the total energy consumption while meeting all the deadlines is constant and equal to total utilization.*

**Proposition 2** [4] *A task assignment that evenly divides the total load among all the processors, if it exists, will minimize the total energy consumption for any number of tasks.*

So, minimizing the makespan will minimize energy consumption especially when full-chip DVFS multiprocessor platforms are considered, the makespan cost function, Eq. (6), will be used as all processors on the chip have to operate at the same frequency which is the maximum processor utilization.

On the other hand, if per-core DVFS multiprocessor platforms are assumed, an energy-aware cost function needs to be proposed. An energy-aware cost function introduced in [3] depends on average utilization of processors, but it does not give an accurate measure for energy consumption. Then, a trade off between average and maximum utilization is introduced.

This paper introduces an energy-aware cost function considering simplified energy model as follows

$$Cost = Sum(U_j^2) / m + Penalty \quad for\ j = 1,2,..,m \quad (8)$$

When applying PSO, the parameters used are the swarm size $k = 100$, No. of iterations=100, $C_1 = C_2 = 2$ [3], and the inertia $W = 1$ that, according to the PSO variant used, may be fixed or may decrease linearly until reaching 0 or it may be then restarted (re-excited) to 1 to decrease linearly again.

## 5. Results and Discussion

The approaches have been implemented using MATLAB[TM]. Utilization matrices have been uniformly generated of light tasks with utilization ranges from 0.05 to 0.25 and medium tasks with utilization ranges from 0.25 to 0.5. The implemented approaches are Min-min, Max-min, PSO with fixed inertia (PSO-fi), PSO with varied inertia (PSO-vi), PSO with re-excited inertia (PSO-re), and our proposed Min-min based PSO approach (PSO-m). Executive experiments have been done to verify the proposed approach.

With relatively small search spaces, all PSO variants show good results with reasonable number of iterations. But, when search spaces grow, so much iterations are needed to get good results using PSO approaches.

PSO variants using variable inertia such as PSO-vi and PSO-re show better performance than PSO with fixed inertia (PSO-fi) with the same number of iterations and the same problem instances.

Figures 2 and 3 below show comparisons among Min-min, Max-min, and PSO variants with 200 iterations for light tasks scheduled on 4 and 10 cores respectively.

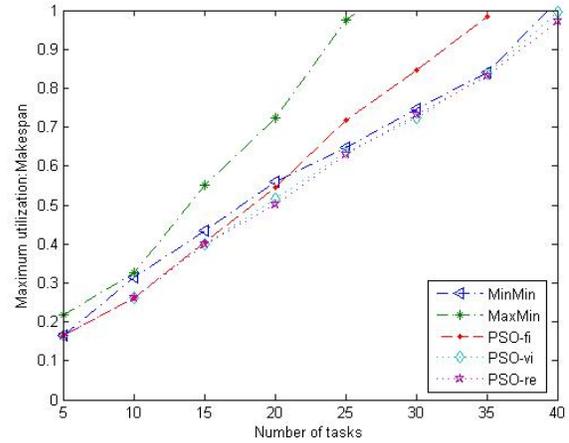

Fig. 2 A comparison of partitioning methods with light tasks partitioned upon 4 processors.

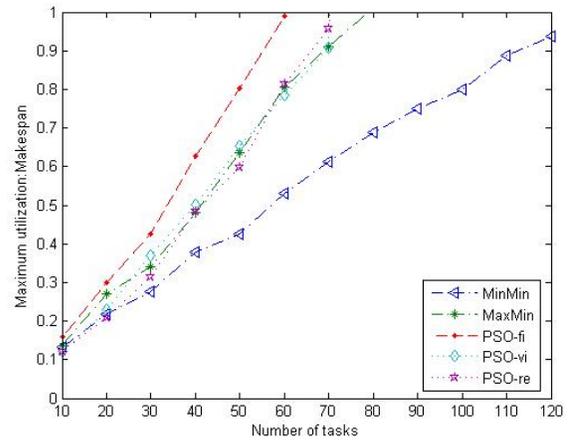

Fig. 3 A comparison of partitioning methods with light tasks partitioned upon 10 processors.

Our proposed approach gives the PSO algorithm a push toward the best solution using a particle (solution) obtained by Min-min. This makes PSO gives better results with reasonable number of iterations. In the worst case, our proposed approach gives Min-min performance if it could not optimize the solution.

Figures 4 and 5 show the performance of our proposed approach with 100 iterations and light tasks assigned to 4 and 10 cores respectively. It is obvious that our proposed approach behaves so better when the search space grows.

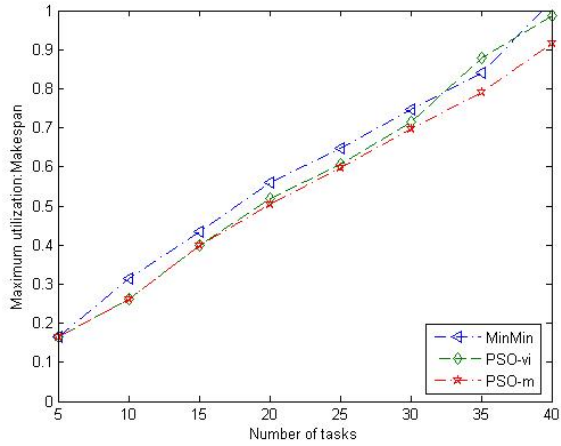

Fig. 4 A comparison among Min-min, PSO-vi, and PSO-m techniques with light tasks partitioned upon 4 processors.

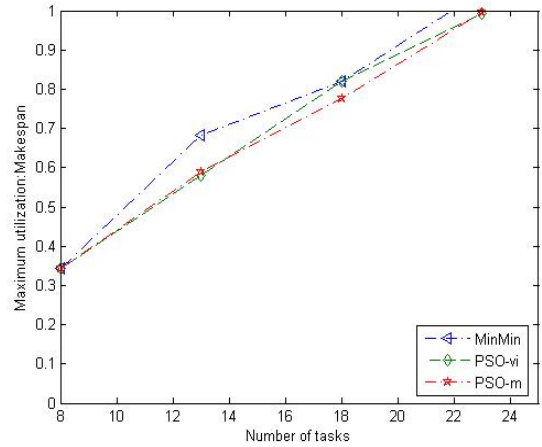

Fig. 6 A comparison among Min-min, PSO-vi, and PSO-m techniques with medium tasks partitioned upon 8 processors.

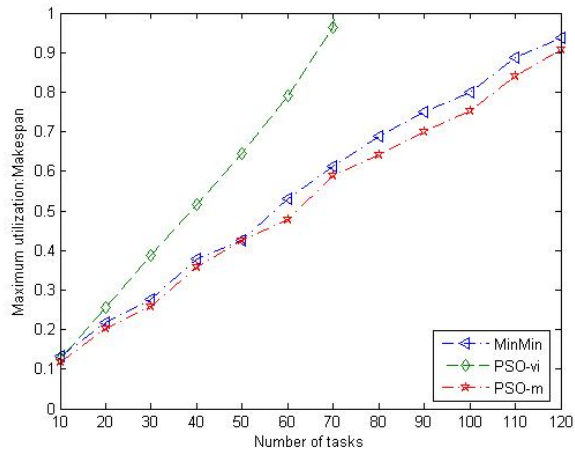

Fig. 5 A comparison among Min-min, PSO-vi, and PSO-m techniques with light tasks partitioned upon 10 processors.

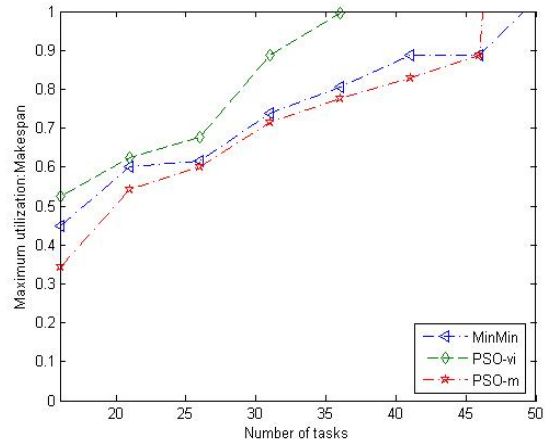

Fig. 7 A comparison among Min-min, PSO-vi, and PSO-m techniques with medium tasks partitioned upon 16 processors.

When medium tasks are used, the proposed approach behaves the same way and shows better performance especially with large search spaces. Figures 6 and 7 show the case when medium tasks are partitioned on 8 and 16 processors respectively.

As mentioned earlier, when full-chip DVFS is considered the makespan cost function is used. If per-core DVFS is considered, the introduced energy-aware cost function, Eq. (8), is taken into account. Figures 8 and 9 show the case of partitioning light tasks on per-core DVFS platforms of 4 and 10 cores respectively.

It is clear that using makespan cost function, Eq. (6), increases the feasibility (schedulability) of the task set more than using Eq. (8) as a cost function which is more energy efficient.

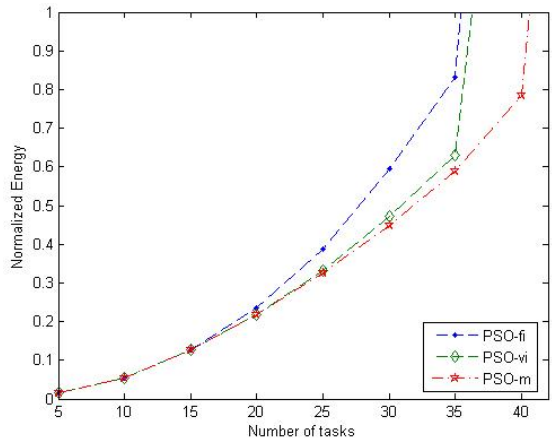

Fig. 8  A comparison among PSO-fi, PSO-vi, and PSO-m techniques with light tasks partitioned upon 4 processors.

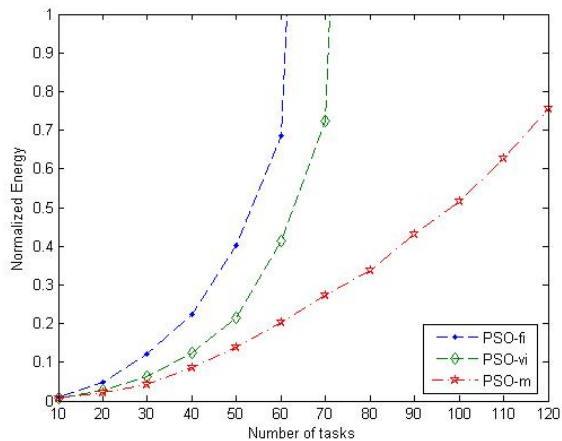

Fig. 9  A comparison among PSO-fi, PSO-vi, and PSO-m techniques with light tasks partitioned upon 10 processors.

It is worth to be noted that another Max-min particle (solution), in addition to Min-min particle, may be added to the population in the initialization step when the task set nature requires that, i.e., when Max-min gives better solutions than Min-min. This occurs when task utilizations are diverse, e.g., when there is a long task in a short-task task set.

## 6. Conclusions

This paper considered the problem of power-aware task partitioning on heterogeneous multiprocessor platforms. The paper proposed a PSO variant based on Min-min that outperformed its counterparts in less number of iterations for the same problem instance. Also, the energy-aware cost function is addressed in this paper and it differentiated between the full-chip and per-core DVFS processors.

As a future work, any verified polynomial-time partitioning technique can be added as a particle to the population in the initialization step to give the PSO algorithm a forward push to get better solutions.